\documentclass[doublecol]{epl2}

\usepackage{graphicx}
\usepackage{subfigure}
\usepackage{amsthm}
\usepackage{amsmath}
\usepackage{amssymb}
\usepackage{verbatim}
\usepackage{dcolumn}
\usepackage{bm}
\usepackage{epsf}
\usepackage{color}
\usepackage[colorlinks=true,citecolor=blue,linkcolor=blue,urlcolor=blue]{hyperref}%
\usepackage{xcolor}
\usepackage{dsfont}

\newcommand{\la}{\left\langle}
\newcommand{\ra}{\right\rangle}
\newcommand{\pd}{\partial}

\newcommand{\e}[1]{\exp{\left(#1\right)}}

\newcommand{\co}[1]{\cos{\left(#1\right)}}
\newcommand{\si}[1]{\sin{\left(#1\right)}}

\newcommand{\bla}{bla\\bla\\bla\\bla\\bla}

\newcommand{\mrm}[1]{\mathrm{#1}}

\title{Nonlinear speed-ups in ultracold quantum gases}

\author{Sebastian Deffner\inst{1,2} \thanks{E-mail: \email{deffner@umbc.edu}}}
\shortauthor{Sebastian Deffner}
\institute{
  \inst{1} Department of Physics, University of Maryland, Baltimore County, Baltimore, MD 21250, USA\\
  \inst{2} Instituto de F\'isica `Gleb Wataghin', Universidade Estadual de Campinas, 13083-859, Campinas, S\~{a}o Paulo, Brazil}

\abstract{Quantum mechanics is an inherently linear theory. However,  collective effects in many body quantum systems can give rise to effectively nonlinear dynamics.  In the present work, we analyze whether and to what extent such nonlinear effects can be exploited to enhance the rate of quantum evolution. To this end, we compute a suitable version of the quantum speed limit for numerical and analytical examples. We find that the quantum speed limit grows with the strength of the nonlinearity, yet it does not trivially scale with the ``degree'' of nonlinearity.  This is numerically demonstrated for the parametric harmonic oscillator obeying Gross-Piteavskii and Kolomeisky dynamics, and analytically for expanding boxes under Gross-Pitaevskii dynamics.
}

\begin{document}

\maketitle


In public perception, the prophesied \emph{quantum advantage} of novel technologies has become almost synonymous with anticipated \emph{quantum speed-ups}.  This impression is driven by quantum computing, which indeed can solve certain problems faster than any classical computer could \cite{Sanders2017}.  At least superficially this expectation seems to be at variance with the so-called \emph{quantum speed limits} (QSL), which are fundamental bounds on the maximal rate with which a quantum system can evolve \cite{Mandelstam1945,Margolus1998}. In fact, diverging QSLs can be interpreted as an herald of classicality \cite{Deffner2017NJP,Poggi2021PRQ}, as they are deeply rooted in more rigorous formulations of Heisenberg's uncertainty relation for energy and time \cite{Heisenberg1927}.

The apparent contradiction quickly dissolves, once one realizes that in the lingo of computer science a ``speed-up'' simply refers to a smaller number of required single gate operations, whereas in quantum physics the QSL refers to the maximal rate with which such a gate operation can be applied \cite{Aifer2022NJP}.  Thus it also becomes rather obvious why so much research activity has been dedicated to the study of QSL in virtually all areas of quantum physics, including, e.g., quantum communication \cite{Bremermann1967,Bekenstein1981,Pendry1983,Caves1994,Lloyd2004,Deffner2020PRR}, quantum computation \cite{Lloyd2000,Mohan2022NJP},  quantum control \cite{Poggi2013EPL,Poggi2016PRA,Kiely2021NJP},  many body physics \cite{Fogarty2020PRL,Puebla2020PRR}, and quantum metrology \cite{Giovannetti2011,Campbell2018}. See somewhat recent reviews on the topic \cite{Frey2016QINP,Deffner2017JPA}.

The original QSLs were formulated for standard quantum mechanics \cite{Pfeifer1995RMP}, whose dynamics is described by the Schr\"odinger equation.  However, over the last decade is has become obvious that there is a variety of ``quantum resources'' that can be employed to speed-up quantum dynamics.  For instance, is has been established that judiciously designed open system dynamics permit environment driven speed-ups \cite{Campo2013,Taddei2013,Deffner2013PRL,Cimmarusti2015,Brody2019,Lan2022NJP}. Similarly, the QSL is larger for non-Hermitian quantum dynamics \cite{Bender2007PRL,Uzdin2012JPA},  for which the quantum states can find literal shortcuts through Hilbert space.

\begin{figure*}[ht]
\includegraphics[width=.31\textwidth]{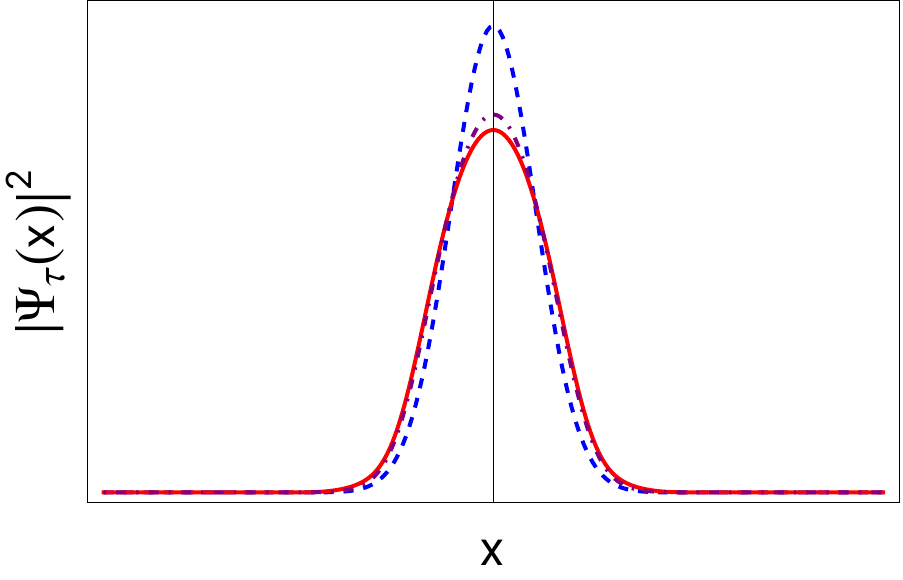}\hfill
\includegraphics[width=.31\textwidth]{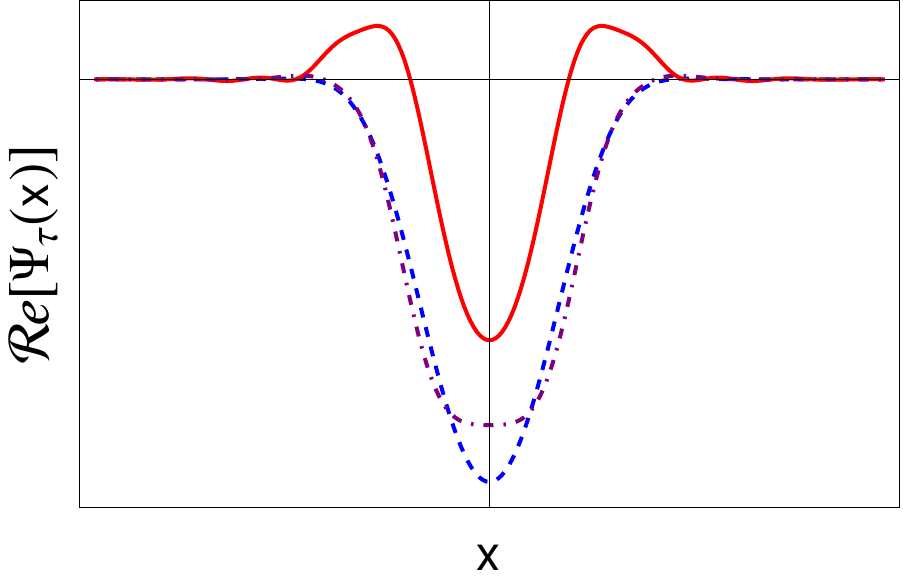}\hfill
\includegraphics[width=.31\textwidth]{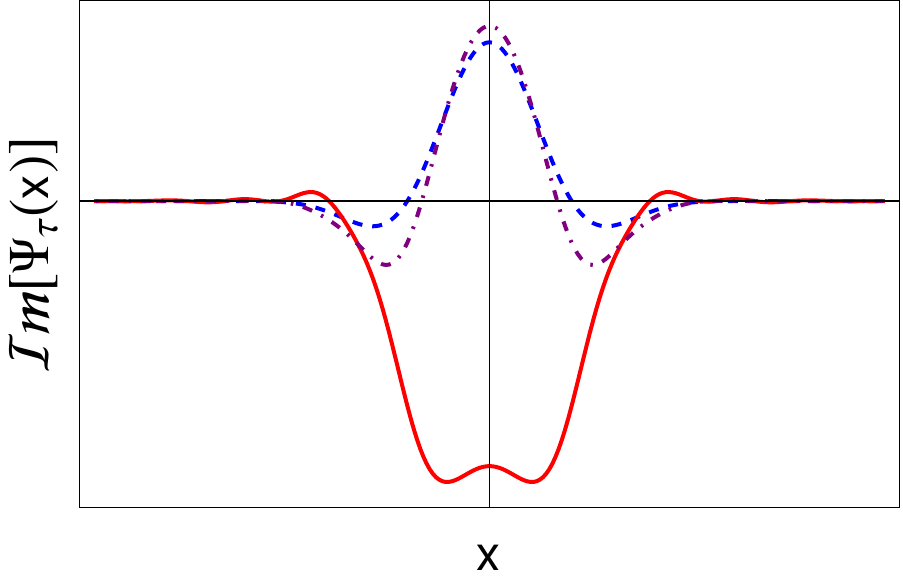}
\caption{\label{fig:harm_wave} From left to right: absolute value, real part, and imaginary part of the solution $\Psi_\tau(x)$ of Eq.~\eqref{eq:non_lin}, for the parametric harmonic oscillator \eqref{eq:U_harm} and $\kappa=0$ (blue, dashed line), $\kappa=5$ (purple, dot-dashed line), and $\kappa=10$ (red, solid line). Other parameters are $\hbar\omega_0=5$, $\hbar\omega_1=1$,  $m=1$, and $\tau=2$}
\end{figure*}

In the present analysis, we will be focusing on another type of quantum dynamics, which has not received much attention in the study of the QSL. Processing information requires some form of interaction between signals. In photonic systems, such interactions can be enabled by nonlinear optical processes \cite{Chang2014}. Nonlinear optics has become almost ubiquitous in fundamental science and technological applications \cite{Rand2010} and can be realized, e.g., with single atoms in cavities \cite{Kirby2015}, in atomic ensembles \cite{Rolston2002}, and through atom-atom interactions \cite{Rajapakse2009}. It is expected that nonlinear systems will be able to outperform linear systems in optical quantum information processing \cite{Chang2014}. The reason is that nonlinear quantum optics can provide both, linear operations such as storage and manipulation of quantum states, as well as nonlinear operations such as the generation of quanta and quantum logic between photonic quantum bits. Recent experimental developments in nonlinear quantum optics hold the promise to enable the development of universal quantum computers supporting conditional quantum logic operations, which are now within technological reach \cite{Chang2014}.  The natural question arises whether nonlinear interactions can be exploited as a quantum resource to enhance the QSL. Some evidence can be found in the literature \cite{Dou2014PRA,Chen2016PRA}, yet a comprehensive analysis within the framework of QSL appears to be lacking.

We show that there is indeed a speed-up of the rate of quantum evolution in nonlinear systems. To this end, we begin with an numerical investigation of a cold bosonic quantum gas in a harmonic trap, which is described by the Gross-Pitaevskii equation \cite{Gross1961,Pitaevskii1961}.  We find that the resulting QSL grows with the strength of the scattering length, meaning that the stronger the nonlinearity the faster the quantum state can evolve. While such numerical evidence is interesting, it is not as insightful as analytical treatments. Therefore, we then solve the Gross-Pitaevskii equation for a 1-dimensional box, whose volume expands at constant rate.  For these dynamics, we obtain an analytical expression for the QSL, and we find our numerical evidence rigorously confirmed.  Thus, as main result, we show that nonlinear quantum effects can quantifiably enhance the rate of quantum evolution. We conclude the analysis by also exploring the QSL for the parametric harmonic oscillator evolving under Kolomeisky dynamics.  Remarkably, we find that the QSL does not trivially grow with the order of the nonlinearity.

\section{Numerical case study: parametric harmonic oscillator}

In the following, we consider quantum dynamics that are described by the Gross-Pitaevskii equation,
\begin{equation}
\label{eq:non_lin}
i\hbar\, \dot{\Psi}_t(x)=\left[-\frac{\hbar^2}{2m}\,\pd_x^2 +U\left(x,t\right)+\kappa\,\left|\Psi_t(x)\right|^2\right]\,\Psi_t(x)\,,
\end{equation}
where, as usual, we denote a derivative with respect to time by a dot. Further, $U(x,t)$ is a time-dependent, external potential and $\kappa$ measures the ``strength'' of the nonlinearity.

Equation~\eqref{eq:non_lin} was first discovered in the description of the propagation of light in nonlinear optical fibers and planar waveguides \cite{Rand2010}, and shortly after in the mean-field description of Bose-Einstein condensates \cite{Gross1961,Pitaevskii1961}.  In cold quantum gases the nonlinearity arises from an effective treatment of the interaction of the bosonic particles, whereas in nonlinear optics $\kappa\,\left|\Psi_t(x)\right|^2$ describes the polarization dependent amplitude in the paraxial field equations. Since its first description,  the nonlinear Schr\"odinger equation \eqref{eq:non_lin} has appeared also in many other areas of physics, such as in hydrodynamics \cite{Nore1993}, in plasma physics \cite{Ruderman2002}, and the propagation of Davydov's alpha-helix solitons, which are responsible for energy transport along molecular chains \cite{Balakrishnan1985}. Thus, Eq.~\eqref{eq:non_lin} has become one of the best studied equations in quantum physics, from a conceptual as well as a numerical point of view \cite{Perez1997,Cerimele2000,Adhikari2002,Bao2003}.

The Gross-Pitaevskii equation \eqref{eq:non_lin} has also found some attention in quantum control \cite{Campo2012SR,Campo2013PRL,Deffner2014PRX,Campo2021PRL}, yet a genuinely nonlinear QSL appears to be lacking in the literature. This may, in parts, be a consequence of the fact that in the derivation of QSL one typically focuses on geometric measures of distinguishability \cite{Pires2016PRX,OConnor2021PRA}.  However,  it also has been argued \cite{Deffner2017NJP,Deffner2020PRR} that in a fully dynamical approach the QSL is simply determined by the rate with which a quantum state evolves, and one can define
\begin{equation}
\label{eq:vQSL}
v_\mrm{QSL}\equiv \int dx \left|\dot{\Psi}_t(x)\right|^2\,.
\end{equation}
It is easy to see that for linear dynamics, $\kappa=0$, we simply have
\begin{equation}
v^0_\mrm{QSL}\equiv v_\mrm{QSL}(\kappa=0)=\la H^2(t)\ra/\hbar^2\,,
\end{equation}
where $H(t)$ is the time-dependent Hamiltonian. Therefore,  Eq.~\eqref{eq:vQSL} can be interpreted as a generalization of the Mandelstam-Tamm bound \cite{Mandelstam1945,Deffner2020PRR}, which remains applicable for any purity preserving quantum dynamics, such as the Gross-Pitaeveskii equation~\eqref{eq:non_lin}.

To build intuition, and to gain some insight into possible nonlinear speed-ups, we now consider the parametric harmonic oscillator, with potential
\begin{equation}
\label{eq:U_harm}
U(x,t)=\frac{1}{2} m \omega_t^2 x^2\,.
\end{equation}
To keep things as simple as possible, we further choose $\omega_t^2$ to be a linear function of time,
\begin{equation}
\label{eq:prot}
\omega^2_t=\omega_0^2-\left(\omega_0^2-\omega_1^2\right) t/ \tau\,,
\end{equation}
for which the linear Schr\"odinger dynamics is analytically solvable \cite{Deffner2008PRE}. For this protocol, we solved Eq.~\eqref{eq:non_lin} numerically for several values of $\kappa$. For each of the realizations, we choose the ground state of the linear problem as initial state, namely
\begin{equation}
\Psi_0(x)=\left(\frac{m \omega_0}{\pi\hbar}\right)^{1/4}\,\e{-\frac{m \omega_0}{2\hbar} x^2}\,.
\end{equation}
The resulting solution is depicted in Fig.~\ref{fig:harm_wave}. We observe that for moderate values of $\kappa$ the solution of the nonlinear dynamics remains close to the linear solution, whereas for large values the dynamics is strongly distinguishable.

\begin{figure*}
\includegraphics[width=.48\textwidth]{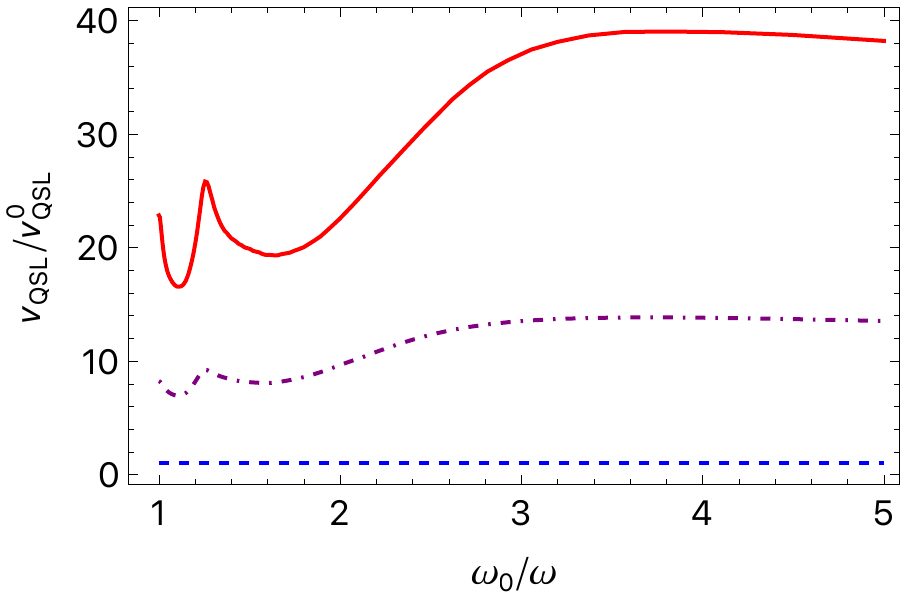}
\includegraphics[width=.48\textwidth]{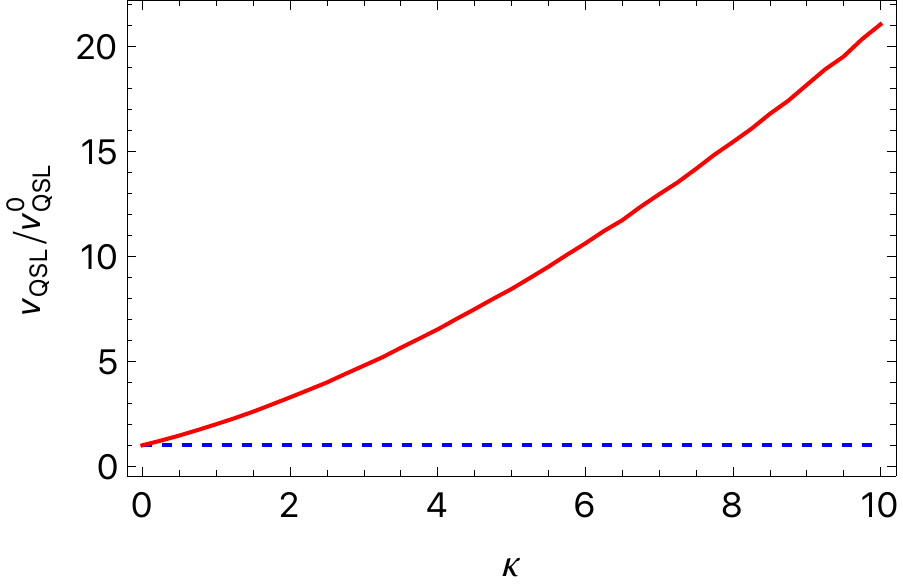}
\caption{\label{fig:harm} Left: QSL \eqref{eq:vQSL} for the parametric harmonic oscillator \eqref{eq:U_harm} for $\kappa=0$ (blue, dashed line), $\kappa=5$ (purple, dot-dashed line), and $\kappa=10$ (red, solid line). Right:  QSL \eqref{eq:vQSL} for the parametric harmonic oscillator \eqref{eq:U_harm} evaluated at $t=\tau/2$ as a function of $\kappa$.  Other parameters are $\hbar\omega_0=5$, $\hbar\omega_1=1$,  $m=1$, and $\tau=2$.}
\end{figure*}

In Fig.~\ref{fig:harm} we plot the corresponding QSL \eqref{eq:vQSL}.  As expected, we observe that the stronger the nonlinearity, i.e., the larger the value of $\kappa$, the faster is the evolution of the quantum state.
While this numerical result is interesting and confirms our intuition, more analytical insight is desirable. In particular, it would be useful to be able to ``design'' optimal nonlinearities to meet, e.g., speed requirements.

\section{Scale-invariant Gross-Pitaevskii dynamics}

Therefore, we now continue with a special class of time-dependent problems.  Consider the scale-invariant, nonlinear Schr\"odinger equation
\begin{equation}
\label{eq:GPE}
i\hbar\, \dot{\Psi}_t(x)=\left[-\frac{\hbar^2}{2m}\,\pd_x^2 +\frac{1}{\lambda_t^2}\,U\left(\frac{x}{\lambda_t}\right)+\frac{\kappa}{\lambda_t}\,\left|\Psi_t(x)\right|^2\right]\,\Psi_t(x),
\end{equation}
where $\lambda_t$ is an external control parameter, such as the volume of an optical trap confining the BEC. Then, if $\lambda_t$ is given for some arbitrary $a$, $b$, and $c$ by $\lambda_t=\sqrt{a t^2+2 b t+c}$, a solution of Eq.~\eqref{eq:GPE} can be written as \cite{Berry1984JPA,Campo2013PRL,Deffner2014PRX}
\begin{equation}
\label{eq:sol}
\Psi_t(x)=\e{-\frac{i}{\hbar} \mu_0 \,\tau(t)}\e{\frac{i m \dot{\lambda}}{2\hbar \lambda}\,x^2}\,\Phi(x,\lambda_t)\,,
\end{equation}
where $\Phi(x,\lambda_t)=\Phi(x/\lambda_t)/\lambda_t^2$ is a scale-invariant solution of the instantaneous Gross-Pitaveskii equation \eqref{eq:GPE},  $\mu_0$ is the initial chemical potential, and $\tau(t)=\int_0^t ds/\lambda_s^2$.  Note that for linear dynamics $\kappa=0$, the chemical potential reduces to the average energy of the initial state, $\epsilon_0$.

It is then a simple exercise to show that the QSL \eqref{eq:vQSL} becomes
\begin{equation}
\label{eq:QSL}
\begin{split}
v_\mrm{QSL}&=\frac{\mu_t^2}{\hbar^2}+\frac{m \mu_t}{\hbar^2}\left(\frac{\dot{\lambda}_t^2}{\lambda_t^2}-\frac{\ddot{\lambda}_t}{\lambda_t}\right)\,\la x^2\ra\\
&+\frac{m^2}{4\hbar^2}\left(\frac{\dot{\lambda}_t^4}{\lambda_t^4}-\frac{2\dot{\lambda}_t^2\ddot{\lambda}_t}{\lambda_t^3}+\frac{\ddot{\lambda}_t^2}{\lambda_t^2}\right)\la x^4\ra+\frac{\dot{\lambda}_t^2}{\hbar^2}\,\la p^2\ra\,,
\end{split}
\end{equation}
where we introduced the instantaneous chemical potential $\mu_t=\mu_0/\lambda_t^2$, and $\la x^n\ra =\int dx\, x^n|\Phi\left(x,\lambda_t\right)|^2$.

\subsection{Example: infinite square well}

As a case study, we continue with an analytically solvable example.  Arguably, the simplest, non-trivial problem in quantum mechanics is a particle trapped in an infinite square well.  The potential simply is
\begin{equation}
\label{eq:U}
U(x)=\begin{cases}
0, \qquad \forall x\in [0,\lambda] \\
\infty, \qquad\text{otherwise}\,.
\end{cases}
\end{equation}
It is easy to see that the one-dimensional box with time-varying length is scale-invariant.  This has been exploited, for instance, in standard thermodynamic problems such as analyzing the validity of the quantum Jarzynski equality \cite{Quan2012PRE,Gong2014PRE}, and to construct explicit expressions for the counterdiabatic field \cite{Campo2012SR,Jarzynski2013PRA,Deffner2014PRX}. 

As a point of reference, and to build intuition we begin with the linear case $\kappa=0$. For completeness, recall that the eigensystem is given by
\begin{equation}
\label{eq:lin}
\psi_n(x,\lambda)=\sqrt{\frac{2}{\lambda}}\,\si{\frac{n\pi}{\lambda}\,x}\quad\text{and}\quad E_n(\lambda)=\frac{\hbar^2 \pi^2}{2 m \lambda^2} \,n^2\,.
\end{equation}
Moreover, we choose a parameterization in which the box is expanded at constant rate $v$, and hence $\lambda_t=\lambda_0+v t$.  Accordingly, the QSL \eqref{eq:QSL} simplifies to read
\begin{equation}
\label{eq:QSL_lin}
v_\mrm{QSL}=\frac{\epsilon_t^2}{\hbar^2}+\frac{m \epsilon_t}{\hbar^2}\,\frac{v^2}{\lambda_t^2}\,\la x^2\ra+\frac{m^2}{4\hbar^2}\,\frac{v^4}{\lambda_t^4}\,\la x^4\ra+\frac{v^2}{\hbar^2}\,\la p^2\ra\,,
\end{equation}
where $\epsilon_t=\epsilon_0/\lambda_t^2$ and $\epsilon_0$ is the initial energy.  We immediately observe that $v_\mrm{QSL}$ is a sum of positive terms, and hence we can continue comparing the nonlinear case to the linear result term-by-term.

The corresponding nonlinear problem can also be solved analytically \cite{Carr2000PRA}, and the stationary states read
\begin{equation}
\label{eq:psi_non}
\psi_n(x,\lambda)=\sqrt{\frac{\nu K(\nu)}{\lambda\,(K(\nu)-E(\nu))}}\, \mrm{sn}\left(\frac{2 n \mrm{K}(\nu)}{\lambda} x \bigg|\nu\right)\,.
\end{equation}
Here $\mrm{sn}(x|\nu)$ is the Jacobian elliptic function \cite{abra}, and $K(\nu)$ and $E(\nu)$ are elliptic integrals. The parameter $\nu$ is directly related to the strength of the nonlinearity $\kappa$. It it is implicitly determined by
\begin{equation}
\label{eq:nu}
 \mrm{K}(\nu)\,\left(K(\nu)-E(\nu)\right)=\frac{m\lambda\,\kappa}{4 n^2\hbar^2}\,,
\end{equation}
which follows from substituting Eq.~\eqref{eq:psi_non} into Eq.~\eqref{eq:GPE} for $U(x)$ as given in Eq.~\eqref{eq:U}. Accordingly, the generalized eigenvalues can be written as
\begin{equation}
\label{eq:mu}
\mu_n(\lambda)=\frac{\hbar^2}{2 m \lambda^2}\,n^2\,(2 \mrm{K}(\nu))^2\,(1+\nu)\,.
\end{equation}
With Eqs.~\eqref{eq:psi_non}--\eqref{eq:mu} the QSL \eqref{eq:QSL_lin} can be computed exactly. However, the resulting expression is rather involved, and hence does not permit to gain much insight. 

Therefore, we now continue to compute corrections to the quantum speed limit up to linear order in the strength of the nonlinearity $\kappa$.  The linear eigensystem \eqref{eq:lin} is recovered for $\nu=0$. Thus, we begin by expanding the left side of Eq.~\eqref{eq:nu} up to linear order in $\nu$ \footnote{See Ref.~\cite{abra} for mathematical details of the elliptic integrals}, from which we obtain
\begin{equation}
\label{eq:nu_a}
\nu\simeq\frac{2 m\lambda\,\kappa}{n^2\hbar^2\pi^2}\,.
\end{equation}
Similarly, the generalized eigenvalues become \cite{Carr2000PRA}
\begin{equation}
\mu_n(\lambda)\simeq\frac{\hbar^2 \pi^2}{2 m \lambda^2} \,n^2 \left(1+ \frac{3 \nu}{2}\right)\,,
\end{equation}
which can be written in terms of the linear energy eigenvalue $E_n(\lambda)$ \eqref{eq:lin} as
\begin{equation}
\mu_n(\lambda)\simeq E_n(\lambda) + \frac{3}{2} \frac{\kappa}{\lambda }\,.
\end{equation}
Hence, for the first term in Eq.~\eqref{eq:QSL_lin} we have already obtained the leading order correction, i.e., ``the nonlinear speed up''.

The other terms in Eq.~\eqref{eq:QSL_lin} require a little more work.  We continue with the second moment of the momentum, which can in fact be computed in closed form.  To further simplify the analysis, we now assume that the system was initially prepared in its ground state, $n=1$. In this case we have
\begin{equation}
\la p^2\ra=\frac{(2 K(\nu))^2}{3 \lambda^2 (K(\nu)-E(\nu))}\,\left[K(\nu)(\nu-1)+E(\nu)(\nu+1)\right]\,,
\end{equation} 
which in leading order of $\nu$ simply becomes
\begin{equation}
\la p^2\ra=\frac{\pi^2}{\lambda^2}+\frac{\pi^2}{32 \lambda^2}\,\nu^2\,.
\end{equation}
In other words, we obtain
\begin{equation}
\la p^2\ra=\la p^2\ra_\mrm{lin}+\frac{m^2 \kappa^2}{8 \hbar^4\pi^2}
\end{equation}
which is again larger than the expression corresponding to linear dynamics,  $\la p^2\ra_\mrm{lin}$, and where we used Eq.~\eqref{eq:nu_a} for $n=1$.

To compute the second and fourth moment of $x$, we now need to expand the stationary states \eqref{eq:psi_non}. For small $\nu$ the Jacobi elliptic function can be approximated by \cite{abra}
\begin{equation}
\mrm{sn}(x|\nu)\simeq \si{x}-\frac{1}{4}\nu \left(x -\si{x}\co{x}\right) \co{x}\,.
\end{equation}
and thus we have
\begin{equation}
\begin{split}
&\psi_n(x,\lambda)\simeq A_n \si{\frac{2 n \mrm{K}(\nu)}{\lambda} x}\\
&\quad-\frac{1}{4}\nu \left(\frac{2 n \mrm{K}(\nu)}{\lambda} x -\si{\frac{2 n \mrm{K}(\nu)}{\lambda} x}\co{\frac{2 n \mrm{K}(\nu)}{\lambda} x}\right)\\
&\quad \times \co{\frac{2 n \mrm{K}(\nu)}{\lambda} x}\,.
\end{split}
\end{equation}
Note that $K(0)=\pi/2$ and hence in leading order we recover the eigenfunctions of the linear case \eqref{eq:lin}. Moreover, $A_n$ is the new normalization coefficient, which can be written in leading order of $\nu$ as
\begin{equation}
\label{eq:A}
A_n\simeq \sqrt{\frac{2}{\lambda}}-\frac{\nu}{8 \sqrt{2\lambda}}\,.
\end{equation}
Accordingly, the second moment of $x$ becomes,
\begin{equation}
\la x^2\ra \simeq\la x^2\ra_\mrm{lin}+\frac{3\lambda^2}{64 \pi^2}\,\nu\,,
\end{equation}
which with Eq.~\eqref{eq:nu_a} can be written as
\begin{equation}
\la x^2\ra \simeq\la x^2\ra_\mrm{lin}+\frac{3m \lambda^3\,\kappa}{32 \hbar^2\pi^4}\,.
\end{equation}
We again observe that the leading order in $\kappa$ is additive and positive. Similarly, we have for the fourth moment
\begin{equation}
\la x^4\ra\simeq\la x^4\ra_\mrm{lin}+\frac{3 (8 \pi^2-15)\lambda^2}{256 \pi^4}\,\nu\,,
\end{equation}
and
\begin{equation}
\la x^4\ra \simeq\la x^4\ra_\mrm{lin}+\frac{3 (8 \pi^2-15)m \lambda^3\,\kappa}{128 \hbar^2 \pi^6}\,.
\end{equation}

\begin{figure*}
\includegraphics[width=.48\textwidth]{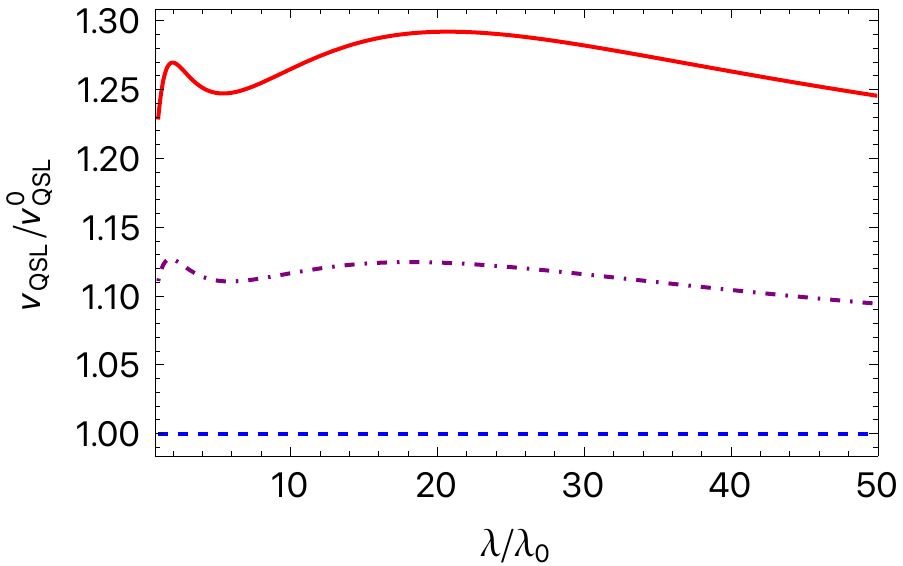}
\includegraphics[width=.48\textwidth]{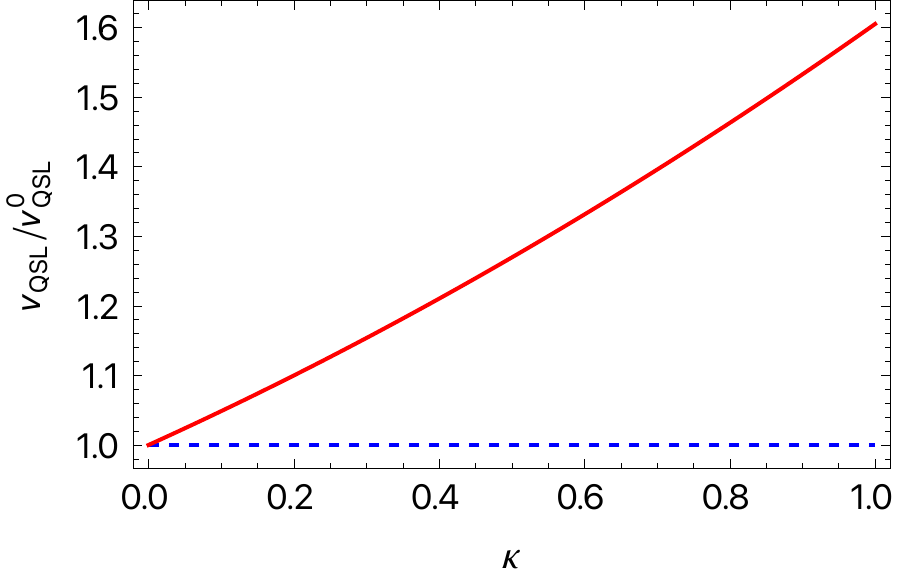}
\caption{\label{fig:box} Left: QSL \eqref{eq:vQSL} for the time-dependent box \eqref{eq:U} for $\kappa=0$ (blue, dashed line), $\kappa=0.25$ (purple, dot-dashed line), and $\kappa=0.5$ (red, solid line). Right:  QSL \eqref{eq:vQSL} for the time-dependent box \eqref{eq:U}  as a function of $\kappa$.  Other parameters are $\hbar=1$, $v=1$,  and $m=1$.}
\end{figure*}

Hence, we immediately conclude that the quantum speed limit \eqref{eq:QSL} for quantum boxes that are prepared in their corresponding ground states, and whose volume is changed at constant rate, is larger for nonlinear dynamics, than for linear dynamics. These findings are illustrated in Fig.~\ref{fig:box}.

In conclusion, the analytically solvable model confirms the numerically found result, namely that nonlinear quantum dynamics supports faster evolution than linear dynamics.  However, from the analytical result we have also obtained exact, closed expressions for $v_\mrm{QSL}$ in leading order of $\kappa$

\begin{figure*}[ht]
\includegraphics[width=.31\textwidth]{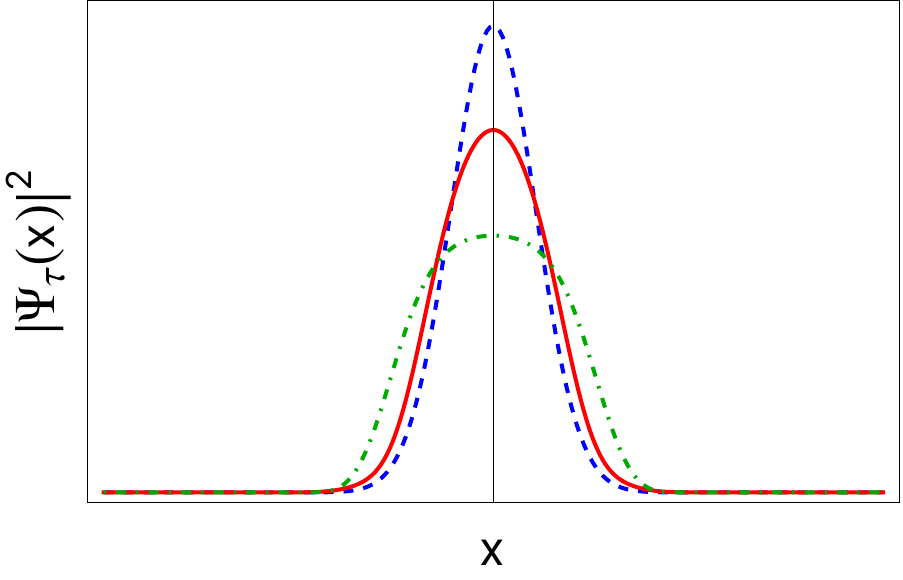}\hfill
\includegraphics[width=.31\textwidth]{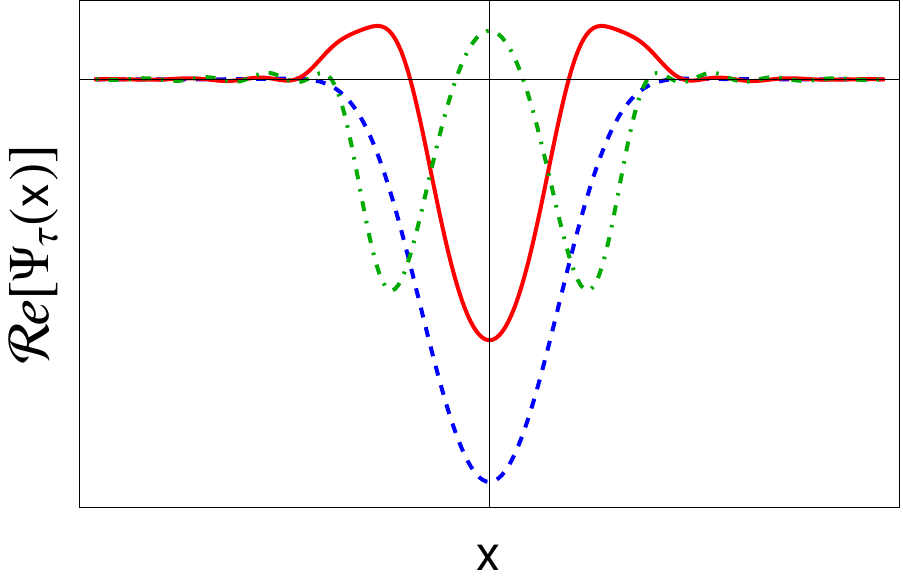}\hfill
\includegraphics[width=.31\textwidth]{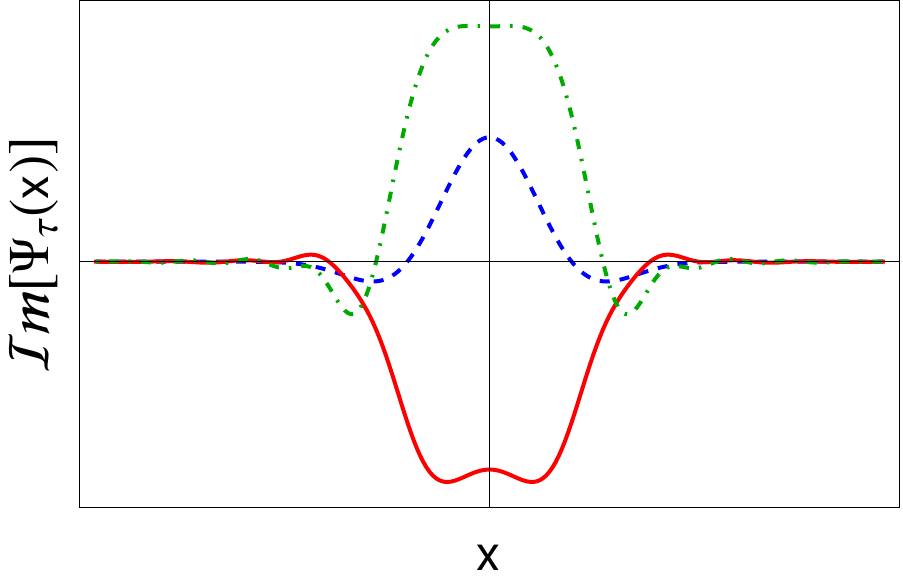}
\caption{\label{fig:harm_wave_2} From left to right: absolute value, real part, and imaginary part of the solution $\Psi_\tau(x)$ for the parametric harmonic oscillator \eqref{eq:U_harm} with linear dynamics (blue, dashed line),  Gross-Piteaveskii dynamics \eqref{eq:non_lin} (red,  solid line), and Kolomeisky dynamic \eqref{eq:kol} (green,  dot-dashed line). Other parameters are $\hbar\omega_0=5$, $\hbar\omega_1=1$,  $m=1$,  $\kappa=10$, and $\tau=2$}
\end{figure*}

\section{Beyond Gross-Pitaevskii dynamics}

In our analysis we have so far restricted ourselves to the 1-dimensional Gross-Pitaevskii equation \eqref{eq:non_lin}. However, it has been noted \cite{Kolomeisky2000PRL} that the description of dilute Bose systems requires a fundamental modification of the dynamics in low dimensions $d\leq 2$.  In fact, such scenarios are better described by
\begin{equation}
\label{eq:kol}
i\hbar\, \dot{\Psi}_t(x)=\left[-\frac{\hbar^2}{2m}\,\pd_x^2 +U\left(x,t\right)+\kappa\,\left|\Psi_t(x)\right|^4\right]\,\Psi_t(x)\,,
\end{equation}
where the nonlinear terms is of fifth order, rather than third in Eq.~\eqref{eq:non_lin}.  For a in-depth discussion of the significance, properties, and experimental relevance of the Kolomeisky equation \eqref{eq:kol} we refer to a recent review article \cite{Sowiski2019RPP}.

For our present purposes the immediate question arises whether making the dynamics ``more nonlinear'' allows for a further speed-up of the evolution.  To systematically answer this question, we solved Eq.~\eqref{eq:kol} numerically for the parametric harmonic oscillator \eqref{eq:U_harm} with the linear protocol \eqref{eq:prot}, and for the same initial state \eqref{eq:lin}.  In Fig.~\ref{fig:harm_wave_2} we depict the result together with linear evolution and the Gross-Pitaveskii dynamics.

We observe that the solution of the Kolomeisky dynamics \eqref{eq:kol} is markedly different from the linear and Gross-Pitaevskii case \eqref{eq:non_lin}. However, from the solution is it not obvious whether the evolution is any faster or slower for the ``more nonlinear dynamics''.

\begin{figure*}
\includegraphics[width=.48\textwidth]{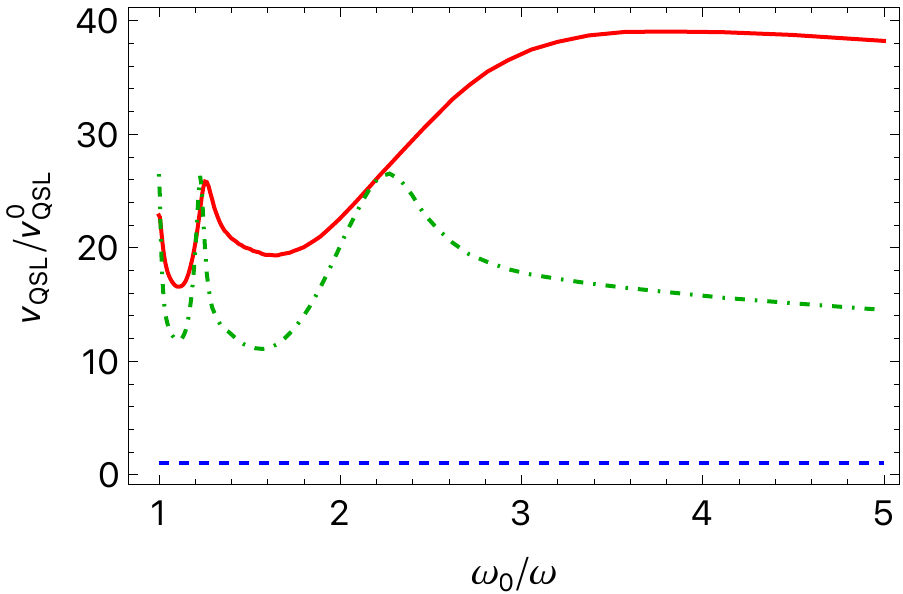}
\includegraphics[width=.48\textwidth]{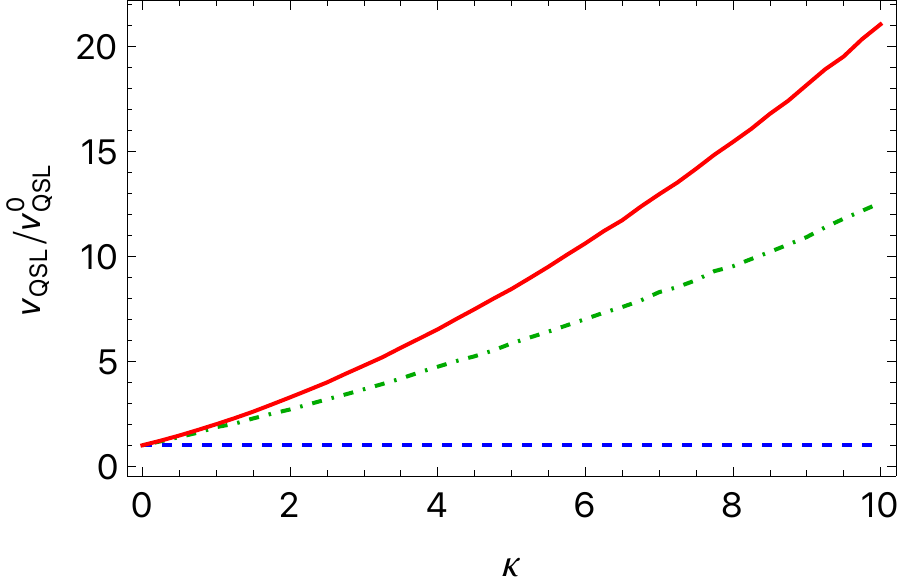}
\caption{\label{fig:harm_2} Left: QSL \eqref{eq:vQSL} for the parametric harmonic oscillator \eqref{eq:U_harm} with linear dynamics (blue, dashed line),  Gross-Piteaveskii dynamics \eqref{eq:non_lin} (red,  solid line), and Kolomeisky dynamic \eqref{eq:kol} (green,  dot-dashed line). Right:  QSL \eqref{eq:vQSL} for the parametric harmonic oscillator \eqref{eq:U_harm} evaluated at $t=\tau/2$ as a function of $\kappa$.  Other parameters are $\hbar\omega_0=5$, $\hbar\omega_1=1$,  $m=1$, and $\tau=2$.}
\end{figure*}

Therefore, we again computed the QSL \eqref{eq:QSL}, which is plotted in Fig.~\ref{fig:harm_2}.
We immediately observe that for the present case, the QSL for Kolomeisky dynamics is almost always smaller than for the Gross-Piteaveskii case. Hence, we conclude that higher order nonlinearities do generally not enhance the rate of quantum evolution. However, now the obvious question arises whether there is an ``optimal'' nonlinearity for which the QSL becomes maximal. Since answering this question will require an in-depth, systemic explorations of physically relevant, nonlinear evolution equations, we leave the answer for future work.

\section{Concluding remarks}

In the present work we have demonstrated, with numerical and analytical examples, that nonlinear quantum dynamics can enhance the rate of quantum evolution.  These findings are not only of academic interest, but they may find application in a variety of practically relevant problems.  For instance, we can foresee that nonlinear quantum speed-ups may be particularly relevant for quantum communication, quantum computation, and quantum thermodynamics.

Nonlinear effects in classical optics have been well-studied in the context of communication \cite{Schneider2004}. However, nonlinear effects in optical fibers are typically very weak \cite{Schneider2004}, since they depend on the intensity of the transmitted signals and the interaction length.  Thus, exploiting nonlinear effects in quantum optics appears more realistic than in classical optics \cite{Boyd2011} . In fact,  the National Institute of Standards and Technology (NIST) has dedicated an entire research program to the exploration and development of nonlinear quantum networks \cite{NIST}.  Our findings clearly show that nonlinear effects will enable faster quantum communication. 

Another area, where nonlinear quantum speed-ups may be relevant, is nonlinear quantum computation. In this framework, unitary gates are replaced with nonlinear quantum operations \cite{Aaronson2005PRSA,Adhikari2013PRL,Holmes2021}. Our results indicate that nonlinear computing not only has a computational advantage, but also that single gate operations can be implemented faster than in linear computers. Hence, the overall processing time can be made shorter, and the quantum computer becomes less susceptible to environmental noise.

Finally, we have recently shown that Bose-Einstein condensation can boost the performance of quantum heat engines \cite{Myers2022NJP}. However, our previous analysis \cite{Myers2022NJP} was restricted to endoreversible cycles, i.e., finite time operation that is still slow enough to allow the working medium to reach a state of local equilibrium. At this point, it is not obvious that the performance boost will persist for engines that operate far from equilibrium. However, our present results suggest that, indeed, the power output may be enhanced by collaborative quantum effects within the working medium, which would constitute a thermodynamic advantage arising from the effectively nonlinear quantum dynamics.  See also Ref.~\cite{Li2018NJP} for similar conclusions.

\acknowledgements{Enlightening discussions with Nathan M. Myers,  Akram Touil,  Maxwell Aifer, and Steve Campbell are gratefully acknowledged. S.D. acknowledges support from the U.S. National Science Foundation under Grant No. DMR-2010127. }

\bibliographystyle{eplbib}
\bibliography{nonlinear_QSL}

\begin{thebibliography}{10}
\expandafter\ifx\csname url\endcsname\relax\def\url#1{\texttt{#1}}\fi

\bibitem{Sanders2017}
\Name{Sanders B.~C.} \Book{How to Build a Quantum Computer} 2399-2891 (IOP
  Publishing) 2017.

\bibitem{Mandelstam1945}
\Name{Mandelstam L. \and Tamm I.} \REVIEW{J. Phys.}{9}{1945}{249}.

\bibitem{Margolus1998}
\Name{Margolus N. \and Levitin L.~B.} \REVIEW{Physica D}{120}{1998}{188}.

\bibitem{Deffner2017NJP}
\Name{Deffner S.} \REVIEW{New J. Phys.}{19}{2017}{103018}.

\bibitem{Poggi2021PRQ}
\Name{Poggi P.~M., Campbell S. \and Deffner S.} \REVIEW{PRX
  Quantum}{2}{2021}{040349}.

\bibitem{Heisenberg1927}
\Name{Heisenberg W.} \REVIEW{Z. f{\"u}r Phys.}{43}{1927}{172}.

\bibitem{Aifer2022NJP}
\Name{Aifer M. \and Deffner S.} \REVIEW{New J. Phys.}{24}{2022}{055002}.

\bibitem{Bremermann1967}
\Name{Bremermann H.~J.} \Book{Quantum noise and information} in proc. of
  \Book{Proceedings of the Fifth Berkeley Symposium on Mathematical Statistics
  and Probability, Volume 4: Biology and Problems of Health} (University of
  California Press, Berkeley, Calif.) 1967 pp. 15--20.

\bibitem{Bekenstein1981}
\Name{Bekenstein J.~D.} \REVIEW{Phys. Rev. Lett.}{46}{1981}{623}.

\bibitem{Pendry1983}
\Name{Pendry J.~B.} \REVIEW{J. Phys. A: Math. Gen.}{16}{1983}{2161}.

\bibitem{Caves1994}
\Name{Caves C.~M. \and Drummond P.~D.} \REVIEW{Rev. Mod. Phys.}{66}{1994}{481}.

\bibitem{Lloyd2004}
\Name{Lloyd S., Giovannetti V. \and Maccone L.} \REVIEW{Phys. Rev.
  Lett.}{93}{2004}{100501}.

\bibitem{Deffner2020PRR}
\Name{Deffner S.} \REVIEW{Phys. Rev. Research}{2}{2020}{013161}.

\bibitem{Lloyd2000}
\Name{Lloyd S.} \REVIEW{Nature}{406}{2000}{1047}.

\bibitem{Mohan2022NJP}
\Name{Mohan B., Das S. \and Pati A.~K.} \REVIEW{New J.
  Phys.}{24}{2022}{065003}.

\bibitem{Poggi2013EPL}
\Name{Poggi P.~M., Lombardo F.~C. \and Wisniacki D.~A.} \REVIEW{{EPL}
  (Europhysics Letters)}{104}{2013}{40005}.

\bibitem{Poggi2016PRA}
\Name{Poggi P.~M.} \REVIEW{Phys. Rev. A}{99}{2019}{042116}.

\bibitem{Kiely2021NJP}
\Name{Kiely A. \and Campbell S.} \REVIEW{New J. Phys.}{23}{2021}{033033}.

\bibitem{Fogarty2020PRL}
\Name{Fogarty T., Deffner S., Busch T. \and Campbell S.} \REVIEW{Phys. Rev.
  Lett.}{124}{2020}{110601}.

\bibitem{Puebla2020PRR}
\Name{Puebla R., Deffner S. \and Campbell S.} \REVIEW{Phys. Rev.
  Research}{2}{2020}{032020}.

\bibitem{Giovannetti2011}
\Name{Giovannetti V., Lloyd S. \and Maccone L.} \REVIEW{Nat.
  Photonics}{5}{2011}{222}.

\bibitem{Campbell2018}
\Name{Campbell S., Genoni M.~G. \and Deffner S.} \REVIEW{Quantum Sci.
  Technol.}{3}{2018}{025002}.

\bibitem{Frey2016QINP}
\Name{Frey M.~R.} \REVIEW{Quantum Inf. Process.}{15}{2016}{3919}.

\bibitem{Deffner2017JPA}
\Name{Deffner S. \and Campbell S.} \REVIEW{J. Phys. A: Math.
  Theor.}{50}{2017}{453001}.

\bibitem{Pfeifer1995RMP}
\Name{Pfeifer P. \and Fr\"ohlich J.} \REVIEW{Rev. Mod. Phys.}{67}{1995}{759}.

\bibitem{Campo2013}
\Name{del Campo A., Egusquiza I.~L., Plenio M.~B. \and Huelga S.~F.}
  \REVIEW{Phys. Rev. Lett.}{110}{2013}{050403}.

\bibitem{Taddei2013}
\Name{Taddei M.~M., Escher B.~M., Davidovich L. \and de~Matos~Filho R.~L.}
  \REVIEW{Phys. Rev. Lett.}{110}{2013}{050402}.

\bibitem{Deffner2013PRL}
\Name{Deffner S. \and Lutz E.} \REVIEW{Phys. Rev. Lett.}{111}{2013}{010402}.

\bibitem{Cimmarusti2015}
\Name{Cimmarusti A.~D., Yan Z., Patterson B.~D., Corcos L.~P., Orozco L.~A.
  \and Deffner S.} \REVIEW{Phys. Rev. Lett.}{114}{2015}{233602}.

\bibitem{Brody2019}
\Name{{Brody} D.~C. \and {Longstaff} B.} \REVIEW{Phys. Rev.
  Research}{1}{2019}{033127}.

\bibitem{Lan2022NJP}
\Name{Lan K., Xie S. \and Cai X.} \REVIEW{New J. Phys.}{24}{2022}{055003}.

\bibitem{Bender2007PRL}
\Name{Bender C.~M., Brody D.~C., Jones H.~F. \and Meister B.~K.} \REVIEW{Phys.
  Rev. Lett.}{98}{2007}{040403}.

\bibitem{Uzdin2012JPA}
\Name{Uzdin R., Günther U., Rahav S. \and Moiseyev N.} \REVIEW{J. Phys. A:
  Math. Theor.}{45}{2012}{415304}.

\bibitem{Chang2014}
\Name{Chang D.~E., Vuleti{\'{c}} V. \and Lukin M.~D.} \REVIEW{Nat.
  Photonics}{8}{2014}{685}.

\bibitem{Rand2010}
\Name{Rand S.} \Book{{Nonlinear and Quantum Optics using the density matrix}}
  (Oxford University Press) 2010.

\bibitem{Kirby2015}
\Name{Kirby B., Hickman G., Pittman T. \and Franson J.} \REVIEW{Optics
  Commun.}{337}{2015}{57 }.

\bibitem{Rolston2002}
\Name{Rolston S.~L. \and Phillips W.~D.} \REVIEW{Nature}{416}{2002}{219}.

\bibitem{Rajapakse2009}
\Name{Rajapakse R.~M., Bragdon T., Rey A.~M., Calarco T. \and Yelin S.~F.}
  \REVIEW{Phys. Rev. A}{80}{2009}{013810}.

\bibitem{Dou2014PRA}
\Name{Dou F.~Q., Fu L.~B. \and Liu J.} \REVIEW{Phys. Rev. A}{89}{2014}{012123}.

\bibitem{Chen2016PRA}
\Name{Chen X., Ban Y. \and Hegerfeldt G.~C.} \REVIEW{Phys. Rev.
  A}{94}{2016}{023624}.

\bibitem{Gross1961}
\Name{Gross E.~P.} \REVIEW{Nuovo Cim.}{20}{1961}{454}.

\bibitem{Pitaevskii1961}
\Name{Pitaevskii L.~P.} \REVIEW{Sov. J. Exp. Theor. Phys.}{13}{1961}{451}.

\bibitem{Nore1993}
\Name{Nore C., Brachet M. \and Fauve S.} \REVIEW{Phys. D Nonlinear
  Phenom.}{65}{1993}{154}.

\bibitem{Ruderman2002}
\Name{Ruderman M.~S.} \REVIEW{Phys. Plasmas}{9}{2002}{2940}.

\bibitem{Balakrishnan1985}
\Name{Balakrishnan R.} \REVIEW{Phys. Rev. A}{32}{1985}{1144}.

\bibitem{Perez1997}
\Name{P\'erez-Garc\'{\i}a V.~M., Michinel H., Cirac J.~I., Lewenstein M. \and
  Zoller P.} \REVIEW{Phys. Rev. A}{56}{1997}{1424}.

\bibitem{Cerimele2000}
\Name{Cerimele M.~M., Chiofalo M.~L., Pistella F., Succi S. \and Tosi M.~P.}
  \REVIEW{Phys. Rev. E}{62}{2000}{1382}.

\bibitem{Adhikari2002}
\Name{Adhikari S.~K. \and Muruganandam P.} \REVIEW{J. Phys. B: At. Mol. Opt.
  Phys.}{35}{2002}{2831}.

\bibitem{Bao2003}
\Name{Bao W., Jaksch D. \and Markowich P.~A.} \REVIEW{J. Comp.
  Phys.}{187}{2003}{318 }.

\bibitem{Campo2012SR}
\Name{Campo A.~d. \and Boshier M.~G.} \REVIEW{Sci. Rep.}{2}{2012}{648}.

\bibitem{Campo2013PRL}
\Name{del Campo A.} \REVIEW{Phys. Rev. Lett.}{111}{2013}{100502}.

\bibitem{Deffner2014PRX}
\Name{Deffner S., Jarzynski C. \and del Campo A.} \REVIEW{Phys. Rev.
  X}{4}{2014}{021013}.

\bibitem{Campo2021PRL}
\Name{del Campo A.} \REVIEW{Phys. Rev. Lett.}{126}{2021}{180603}.

\bibitem{Pires2016PRX}
\Name{Pires D.~P., Cianciaruso M., C\'eleri L.~C., Adesso G. \and Soares-Pinto
  D.~O.} \REVIEW{Phys. Rev. X}{6}{2016}{021031}.

\bibitem{OConnor2021PRA}
\Name{O'Connor E., Guarnieri G. \and Campbell S.} \REVIEW{Phys. Rev.
  A}{103}{2021}{022210}.

\bibitem{Deffner2008PRE}
\Name{Deffner S. \and Lutz E.} \REVIEW{Phys. Rev. E}{77}{2008}{021128}.

\bibitem{Berry1984JPA}
\Name{Berry M.~V. \and Klein G.} \REVIEW{J. Phys. A: Math.
  Gen.}{17}{1984}{1805}.

\bibitem{Quan2012PRE}
\Name{Quan H.~T. \and Jarzynski C.} \REVIEW{Phys. Rev. E}{85}{2012}{031102}.

\bibitem{Gong2014PRE}
\Name{Gong Z., Deffner S. \and Quan H.~T.} \REVIEW{Phys. Rev.
  E}{90}{2014}{062121}.

\bibitem{Jarzynski2013PRA}
\Name{Jarzynski C.} \REVIEW{Phys. Rev. A}{88}{2013}{040101}.

\bibitem{Carr2000PRA}
\Name{Carr L.~D., Clark C.~W. \and Reinhardt W.~P.} \REVIEW{Phys. Rev.
  A}{62}{2000}{063610}.

\bibitem{abra}
\Name{Abramowitz M. \and Stegun I.} \Book{Handbook of Mathematical Functions
  with Formulas, Graphs, and Mathematical Tables} (United States Department of
  Commerce, National Bureau of Standards (NBS)) 1964.

\bibitem{Kolomeisky2000PRL}
\Name{Kolomeisky E.~B., Newman T.~J., Straley J.~P. \and Qi X.} \REVIEW{Phys.
  Rev. Lett.}{85}{2000}{1146}.

\bibitem{Sowiski2019RPP}
\Name{Sowi{\'{n}}ski T. \and Garc{\'{\i}}a-March M.~{\'{A}}.} \REVIEW{Rep.
  Prog. Phys.}{82}{2019}{104401}.

\bibitem{Schneider2004}
\Name{Schneider T.} \Book{Nonlinear Optics in Telecommunications} (Springer
  Berlin Heidelberg, Berlin, Heidelberg) 2004.

\bibitem{Boyd2011}
\Name{Boyd R.~W., Shin H., Malik M., O'Sullivan C., Chan K. W.~C., Chang H.~J.,
  Gauthier D.~J., Jha A., Leach J., Murugkar S. \and Rodenburg B.}
  \Book{Applications of nonlinear optics in quantum imaging and quantum
  communication} in proc. of \Book{Nonlinear Optics} (Optica Publishing Group)
  2011 p. NWC2.

\bibitem{NIST}
\Book{Nonlinear optics for quantum information and networking}
  https://www.nist.gov/programs-projects/nonlinear-optics-quantum-information-and-networking
  "National Institute for Standards and Technologies (NIST)" (2020).

\bibitem{Aaronson2005PRSA}
\Name{Aaronson S.} \REVIEW{Proc. R. Soc. A}{461}{2005}{3473}.

\bibitem{Adhikari2013PRL}
\Name{Adhikari P., Hafezi M. \and Taylor J.~M.} \REVIEW{Phys. Rev.
  Lett.}{110}{2013}{060503}.

\bibitem{Holmes2021}
\Name{Holmes Z., Coble N., Sornborger A.~T. \and Suba{\c{s}}{\i} Y.}
  \REVIEW{arXiv preprint arXiv:2112.12307}{}{2021}{}.

\bibitem{Myers2022NJP}
\Name{Myers N.~M., Pe{\~{n}}a F.~J., Negrete O., Vargas P., Chiara G.~D. \and
  Deffner S.} \REVIEW{New J. Phys.}{24}{2022}{025001}.

\bibitem{Li2018NJP}
\Name{Li J., Fogarty T., Campbell S., Chen X. \and Busch T.} \REVIEW{New J.
  Phys.}{20}{2018}{015005}.

\end{thebibliography}

\end{document}